\documentclass[%
 reprint,
 amsmath,amssymb,
 aps,
]{revtex4-2}

\usepackage{graphicx}
\usepackage{dcolumn}
\usepackage{bm}
\usepackage{amsmath}
\usepackage{empheq}
\usepackage{tikz}
\usetikzlibrary{graphs,quotes}
\usetikzlibrary{math}
\usetikzlibrary{calc}

\begin{document}


\title{Explicit mutual information for simple networks and neurons with lognormal
activities}

\author{Maurycy Chwi{\l}ka, Jan Karbowski$^{*}$}

\affiliation{\it Department of Mathematics, Informatics, and Mechanics;
  Institute of Applied Mathematics and Mechanics,
University of Warsaw, ul. Banacha 2, 02-097 Warsaw, Poland}




\begin{abstract}
Networks with stochastic variables described by heavy tailed lognormal distribution
are ubiquitous in nature, and hence they deserve an exact information-theoretic
characterization. We derive analytical formulas for mutual information between
elements of different networks with correlated lognormally distributed activities. 
In a special case, we find an explicit expression for mutual information between
neurons when neural activities and synaptic weights are lognormally distributed,
as suggested by experimental data. Comparison of this expression with the case
when these two variables have short tails, reveals that mutual information with
heavy tails for neurons and synapses is generally larger and can diverge for
some finite variances in presynaptic firing rates and synaptic weights.
This result suggests that evolution might prefer brains with
heterogeneous dynamics to optimize information processing.

\begin{description}
\item[Keywords]
  mutual information, lognormal distribution, neuronal information, networks.
\end{description}
\end{abstract}


\maketitle


\section{\label{sec:level1}Introduction}

Lognormal distribution is arguably one of the most common continuous probability distributions
describing naturally occurring phenomena in nature \cite{limpert,crow}: from environment
and geology \cite{limpert,sornette}, to stock market fluctuations and finance
\cite{black,bouchaud}, to brain activity \cite{song,hromadka,buzsaki}. Some other
examples are included in Refs. \cite{tang,corral,sobkowicz}. Lognormal distribution
has skewed shape and is characterized by a long tail \cite{crow}. This means that
the likelihood of huge size stochastic events with this distribution is small,
but still significantly higher than for those described by short-tailed distributions,
e.g., by Gaussian density.

Two or more stochastic variables can be correlated, and one of the most useful
tools describing such correlation is mutual information (MI). MI is an
information-theoretic concept \cite{cover,shannon} that measures average number of
bits we gain about behavior of one variable by observing variability of another,
correlated variable. Obviously, MI has a wide applicability in many different
areas of science, from information science \cite{shannon,cover} and rapidly
developing machine and deep learning \cite{lin,gabrie}, through different branches
of physics (as diverse as material science, stochastic thermodynamics, and
cosmology) \cite{detomasi,iizuka,golub,seifert}, to molecular biology
\cite{wang,cheong} and neuroscience \cite{rieke}. Despite this huge popularity
in using MI, it is important to stress that its exact analytical form for nontrivial
continuous distributions is known only in few cases \cite{darbellay}, which are
however unexplored. It seems that in practical applications only MI for short tailed
Gaussian distribution is used, e.g., \cite{cover,cheong,rieke,paninski}. 
On the other hand, exact form of MI for heavy tailed lognormal distribution is
rarely mentioned and virtually not used (but see a recent notable exception
in \cite{granero}). Thus clearly, there is a need for exploration of general
properties of MI for lognormal variables in different settings.

In this paper, we present and use the general formula for MI between vectors of
random variables with lognormal distributions to obtain explicit analytical expressions
for MI in specific networks with simple topology, which can appear in many applications.
Furthermore, we apply this formalism to neuroscience, and derive analytical MI for
information transfer between neurons in the neural networks relevant for brain cerebral
cortex, i.e., when a neuron receives many correlated synaptic inputs with heavy tails.

\section{\label{sec:level2}Preliminaries}

\subsection{\label{sec:levell2} Lognormal distribution}

Let ${\bf z} = (z_1, z_2, \dots, z_N)$ be a $N$ dimensional Gaussian random variable
with mean vector ${\bf\mu}=(\mu_{1},...,\mu_{N})$ and covariance matrix $\Sigma$,
which is positive definite (symmetric with real and positive eigenvalues).
We define a new multidimensional random variable ${\bf x}=(x_1, x_2, \dots, x_N)$,
such that $x_{i} = e^{z_{i}}$ for every $i=1,...,N$. Then ${\bf x}$ is lognormally
distributed and has the following probability density $\rho({\bf x})$

\begin{widetext}
\begin{eqnarray}
\rho(\bm{x}) = \frac{1}
    {\sqrt{(2\pi)^N\det{(\Sigma)}}\prod_{i=1}^{N}x_{i}}
        \exp{(-\frac{1}{2}(\ln{(\bm{x})}-\bm{\mu})^T\Sigma^{-1}
        (\ln{(\bm{x})}-\bm{\mu}))}, 
\end{eqnarray}
\end{widetext}
with $\langle \ln(x_{i})\rangle= \mu_{i}$, and the covariance matrix 
$\Sigma_{ij}= \langle \ln(x_{i})\ln(x_{j})\rangle
- \langle \ln(x_{i})\rangle\langle \ln(x_{j})\rangle$,
where $\langle ...\rangle$
denotes averaging with respect to the density in Eq. (1). 
We denote the variance of $\ln x_{i}$ as $\Sigma_{ii}= \langle \ln^{2}(x_{i})\rangle
- \langle \ln(x_{i})\rangle^{2} \equiv \sigma_{i}^{2}$.
For $N= 1$, the density in Eq. (1) reduces to
$ \rho(x) = e^{-(\ln{(x)} - \mu)^{2}/2\sigma^2}/(\sqrt{2\pi\sigma^2}x)$.

The major moments of the lognormal distribution, including variance
$Var(x_{i})= \langle x_{i}^{2}\rangle - \langle x_{i}\rangle^{2}$, and
covariance  $Cov(x_{i},x_{j})= \langle x_{i}x_{j}\rangle - \langle x_{i}\rangle
  \langle x_{j}\rangle$, are related to the parameters
$\mu_{i}$ and $\Sigma_{ij}$ as \cite{crow}:
\begin{eqnarray}
 \langle x_{i}\rangle = e^{\mu_{i} + \frac{1}{2}\sigma_{i}^{2}},   \nonumber \\
  Var(x_{i}) = (e^{\sigma_{i}^{2}} - 1)\langle x_{i}\rangle^{2},     \nonumber  \\
  Cov(x_{i},x_{j}) =  (e^{\Sigma_{ij}} - 1)\langle x_{i}\rangle\langle x_{j}\rangle.
\end{eqnarray}

\subsection{\label{sec:levell1} Definition of mutual information}

Mutual information MI between two lognormal random variables X and Y is
defined as \cite{cover}

\begin{equation}\label{eq:mi}
\mbox{MI}(X, Y) = h(X) + h(Y) - h(X, Y),
\end{equation}
where $h(X)$, and $h(Y)$ are differential entropies for X and Y variables,
i.e.

\[
h(X) = -\frac{1}{\ln{2}}\int \rho(\bm{x})\ln{(\rho(\bm{x}))}\,d\bm{x},
\]

and $h(X,Y)$ is the joint differential entropy given by

\[
h(X, Y) = -\frac{1}{\ln{2}}\int \rho(\bm{x}, \bm{y})\ln{(\rho(\bm{x}, 
\bm{y}))}\,d\bm{x}\,d\bm{y},
\]
where $\rho({\bf x})$ is as in Eq. (1) and $\rho({\bf x,y})$ is the joint
probability density of ${\bf x}$ and ${\bf y}$ (see below).
MI(X,Y) quantifies the amount of information we gain about X by observing Y, or
vice versa.

\subsection{\label{sec:levell1} Mutual information for lognormal stochastic vectors}


In this section we present a nonstandard derivation of MI for two random vectors
with lognormal distributions.

First we determine entropy of $N$ dimensional variable ${\bf x}$. To achieve this,
we need the average of the argument of $\exp$ in Eq. (1). We have
$\langle [\ln(\bm{x})-\mu]^T\Sigma^{-1} [\ln(\bm{x})-\mu)] \rangle =
\sum_{ij} (\Sigma^{-1})_{ij}
\langle [\ln(x_{i})-\mu_{i}][\ln(x_{j})-\mu_{j}] \rangle =
\sum_{ij} (\Sigma^{-1})_{ij}\Sigma_{ji}= \sum_{ij} (\Sigma^{-1}\Sigma)_{ii}= N$,
where we used the definition of covariance matrix appearing after Eq. (1).
After some additional straightforward steps we obtain the formula
for entropy of $N$ dimensional lognormal random variable ${\bf x}$:

\begin{empheq}{align}
  h({\bf x}) & = -\frac{1}{\ln{2}}
  \langle \ln{(\rho(\bm{x}))} \rangle  \nonumber  \\
  & = \frac{1}{\ln{2}}\Big[\frac{1}{2}\ln{((2\pi 
    e)^N\det{(\Sigma)})}
    + \sum_{i=1}^N \langle \ln{(x_i)}\rangle \Big]\label{eq:ent}.
\end{empheq}

Now, we partition our dimensionality $N$ into two parts, $N= n+k$,
and define two new multidimensional lognormal random variables $X$
and $Y$ such that ${\bf x}= (X,Y)$, where 
$X = (x_1, x_2, \dots, x_n)$ and $Y = (y_{1}, y_{2}, \dots, y_{k})$,
with $y_{i}\equiv x_{n+i}$.
The lognormal distribution of $X$ has the parameters ${\bf \mu_{X}}$ and $\Sigma_X$,
defined as ${\bf \mu_X}= (\langle\ln x_{1}\rangle,...,\langle\ln x_{n}\rangle)$ and
$n\times n$ matrix $(\Sigma_X)_{ij}= \langle \ln(x_{i})\ln(x_{j})\rangle
- \langle \ln(x_{i})\rangle\langle \ln(x_{j})\rangle$.
The lognormal distribution of $Y$ has the parameters ${\bf \mu_Y}$ and $\Sigma_Y$,
defined as ${\bf \mu_Y}= (\langle\ln y_{1}\rangle,...,\langle\ln y_{k}\rangle)$ and
$k\times k$ matrix $(\Sigma_Y)_{ij}= \langle \ln(y_{i})\ln(y_{j})\rangle
- \langle \ln(y_{i})\rangle\langle \ln(y_{j})\rangle$.
$N$ dimensional variable ${\bf x}$ is lognormally distributed with vector
of means $({\bf \mu_X, \mu_Y})$ and $N\times N$ global covariance matrix
$\Sigma_{XY}$ dependent on $\Sigma_{X}$ and $\Sigma_{Y}$ matrices through the
block matrix

\begin{eqnarray}
\Sigma_{XY} = 
    \begin{bmatrix}
        \Sigma_{X} & Cov_{XY}  \\
        Cov_{XY}^{T} & \Sigma_{Y}
    \end{bmatrix},
\end{eqnarray}
where $Cov_{XY}$ is $n\times k$ covariance matrix between variables
$\ln X$ and $\ln Y$, i.e., $Cov_{XY}\equiv Cov(\ln X, \ln Y)$, and
$(Cov_{XY})_{ij}= \langle \ln(x_{i})\ln(y_{j})\rangle
- \langle \ln(x_{i})\rangle\langle \ln(y_{j})\rangle$.

Consequently, from Eq. (\ref{eq:ent}) it follows that we can write entropies
for each of the three lognormal variables $X$, $Y$, and ${\bf x}$ as

\begin{gather*}
        h(X) = \frac{1}{\ln{2}}\Big[\frac{1}{2}\ln{((2\pi 
    e)^n\det{(\Sigma_X)})}
    + \sum_{i=1}^n \langle \ln{(x_i)}\rangle\Big]\\
    h(Y) = \frac{1}{\ln{2}}\Big[\frac{1}{2}\ln{((2\pi 
    e)^k\det{(\Sigma_Y)})}
    + \sum_{i=n+1}^{N} \langle \ln{(x_i)}\rangle\Big]\\
    h(X, Y) = \frac{1}{\ln{2}}\Big[\frac{1}{2}\ln{((2\pi 
        e)^{N}\det{(\Sigma_{XY})})} + \sum_{i=1}^{N}
      \langle \ln{(x_i)}\rangle\Big].
\end{gather*}
Now, using Eq. \ref{eq:mi}, we can observe that
\begin{eqnarray}\label{eq:mutualinformation}
     \mbox{MI}(X, Y) = \frac{1}{2\ln2}\ln{\Big(\frac{\det{(\Sigma_X)}
    \det{(\Sigma_Y)}}{\det{(\Sigma_{XY})}}\Big)}.
\end{eqnarray}
This is the general formula for mutual information of two random
multidimensional lognormal variables, and it depends only
on the matrices of covariance of corresponding Gaussian random
vectors with appropriate parameters - both joint and
marginal distributions. Thus MI for lognormal variables is formally
the same as MI for the underlying Gaussian variables \cite{cover},
which is a consequence of the invariance of MI with respect to any bijective
(one-to-one) transformation of the variables involved (see \cite{granero}
for this type of approach).

The formula (6) can be rewritten in an
equivalent form if we use a Schur's decomposition for the determinant of
the block matrix $\Sigma_{XY}$, namely \cite{horn,silvester}

\begin{eqnarray}
\det\Sigma_{XY}=
 \det(\Sigma_{X}) \det(\Sigma_{Y}-Cov_{XY}^{T}\Sigma_{X}^{-1}Cov_{XY})  \nonumber \\
 = \det(\Sigma_{Y}) \det(\Sigma_{X}-Cov_{XY}\Sigma_{Y}^{-1}Cov_{XY}^{T}).  
\end{eqnarray}
Then the mutual information is

\begin{eqnarray}
  \mbox{MI}(X,Y) = \frac{1}{\ln4}\ln{\Big(\frac{\det{(\Sigma_Y)}}
    {\det(\Sigma_{Y}-Cov_{XY}^{T}\Sigma_{X}^{-1}Cov_{XY})}
    \Big)}    \nonumber  \\
  = \frac{1}{\ln4}\ln{\Big(\frac{\det{(\Sigma_X)}}
    {\det(\Sigma_{X}-Cov_{XY}\Sigma_{Y}^{-1}Cov_{XY}^{T})} \Big)}
  \nonumber  \\
  = \frac{1}{\ln4}\ln{\Big(\frac{1}
    {\det(I - \Sigma_{X}^{-1}Cov_{XY}\Sigma_{Y}^{-1}Cov_{XY}^{T})} \Big)},  
\end{eqnarray}
where $I$ is the identity matrix, and the last equality follows from a known
identity, $\det(AB)= \det(A)\det(B)$, valid for two arbitrary square matrices
$A$ and $B$. Eqs. (6) and (8) are the first major result of this paper, which
allow us to obtain exact analytical expressions for mutual information in some
specific cases. Note that MI(X,Y) in Eq. (8) is nonzero only if there are
correlations between X and Y, i.e., when the covariance matrix $Cov_{XY}$ is
nonzero. Depending on the structure of matrices $\Sigma_{X}$, $\Sigma_{Y}$, and
$Cov_{XY}$, one can use either of the three expressions in Eq. (8), whichever
is easier, for finding MI(X,Y).

\section{\label{sec:level3}  Analytical results for MI in specific networks}

In this section we consider several specific examples of networks with simple
topology, and give explicit formulas for MI in each case. These networks are
relatively simple to analyze, and are particular instances of the so-called
bipartite networks with correlations that frequently appear in biological and
social contexts \cite{milo,albert,newman,dorogovtsev,peltomaki}.
Elements of the networks below are illustrated by graphs with nodes,
which are connected by arrows. An arrow between the nodes indicates
a correlation among particular elements with random lognormal
activities.

\subsubsection{Case $n =k =1$.}
In this case, $X= x_{1}$ and $Y=y_{1}= x_{2}$, and the global covariance
matrix appearing in Eq. (5) is $2\times 2$, and has the following form

\begin{eqnarray}
\Sigma_{XY} = 
    \begin{bmatrix}
        \sigma_X^2 & Cov_{XY} \\
        Cov_{XY} & \sigma_Y^2
    \end{bmatrix},
\end{eqnarray}
where $\sigma_{X}^{2}= Var(\ln X)= \langle \ln^{2}(X)\rangle
- \langle \ln(X)\rangle^{2}$, and
$\sigma_{Y}^{2}= Var(\ln Y)= \langle \ln^{2}(Y)\rangle
- \langle \ln(Y)\rangle^{2}$. As a result
\begin{eqnarray}
  \mbox{MI}(X,Y) & = \frac{1}{2\ln{2}}\ln{\Big(\frac{\sigma_X^2\sigma_Y^2}
        {\sigma_X^2\sigma_Y^2 - Cov_{XY}^2} \Big)}  \nonumber  \\
    = \frac{1}{2\ln{2}}\ln{\Big(\frac{1}
        {1 - c^2}\Big)},
\end{eqnarray}    
where $c$ denotes correlation coefficient between variables
$\ln(X)$ and $\ln(Y)$, defined as
$c= Cov_{XY}/\sigma_{X}\sigma_{Y}$. The coefficient $c$
can be expressed by covariance of $X$ and $Y$, $Cov(X,Y)$ and their
variances $Var(X), Var(Y)$ as

\begin{eqnarray}
  c = \frac{\ln[ 1 + \frac{Cov(X,Y)}{\langle X\rangle\langle Y\rangle} ]}
  {\sqrt{\ln[ 1 + \frac{Var(X)}{\langle X\rangle^{2}}]
      \ln[ 1 + \frac{Var(Y)}{\langle Y\rangle^{2}}] } },
\end{eqnarray}
which follows from Eq. (2). Because of the Cauchy-Schwartz inequality
for covariance and variance, we have $Cov(X,Y) \le \sqrt{Var(X)Var(Y)}$,
and this implies that $c$ is bounded from above by $c_{0}$, which is

\begin{eqnarray}
  c_{0} = \frac{\ln[ 1 + \frac{\sqrt{Var(X)Var(Y)}}{\langle X\rangle\langle Y\rangle} ]}
  {\sqrt{\ln[ 1 + \frac{Var(X)}{\langle X\rangle^{2}}]
      \ln[ 1 + \frac{Var(Y)}{\langle Y\rangle^{2}}] } }.
\end{eqnarray}
Consequently, we have an upper bound on MI(X,Y) in this case as 

\begin{eqnarray}
  \mbox{MI}(X,Y) &  \le  \frac{1}{2\ln{2}}\ln\Big(\frac{1}
        {1 - c_{0}^2}\Big).
\end{eqnarray}    
This inequality may be useful in cases when we do not know covariances but
we do know variances of original data.

\subsubsection{Case $n =k =2$.}

In this case, $X= (x_{1},x_{2})$ and $Y= (y_{1},y_{2})$, and the covariance matrix
$\Sigma_{XY}$ appearing in Eq. (5) is $4\times 4$. We choose the components of
$\Sigma_{XY}$ with some degree of symmetry:
$(\Sigma_{X})_{ij}= \sigma_{x}^{2}\delta_{ij} + \gamma_{x}(1-\delta_{ij})$,
$(\Sigma_{Y})_{ij}= \sigma_{y}^{2}\delta_{ij} + \gamma_{y}(1-\delta_{ij})$,
and $(Cov_{XY})_{ij}= a\delta_{ij} + b(1-\delta_{ij})$, where $i,j= 1,2$.
Here, all four variables $x_{1}, x_{2}, y_{1}, y_{2}$ are mutually correlated
(see the graph below).

\begin{center}

\begin{tikzpicture}

\draw (0,0) node[draw, circle] (I) {$X_1$}; 
\draw (1.5,0) node[draw, circle] (J) {$X_2$};

\draw (0,-1.5) node[draw, circle] (U) {$Y_1$}; 
\draw (1.5,-1.5) node[draw, circle] (M) {$Y_2$}; 
 
\draw[<->] (I) to (J);
\draw[<->] (U) to (M);
\draw[<->] (I) to (U);
\draw[<->] (J) to (M);
\draw[<->] (I) to (M);
\draw[<->] (U) to (J);

\end{tikzpicture}

\end{center}

Thus we have $\det(\Sigma_{Y})= \sigma_{y}^{4} - \gamma_{y}^{2}$  and
$\det(\Sigma_{X})= \sigma_{x}^{4} - \gamma_{x}^{2}$. Additionally,
we obtain

\begin{eqnarray*}
\det[\Sigma_{Y}-Cov_{XY}^{T}\Sigma_{X}^{-1}Cov_{XY}]
       \nonumber  \\
       = \frac{[(\sigma_{x}^{2} -\gamma_{x})(\sigma_{y}^{2} -\gamma_{y}) - (a-b)^{2}]}
    {\det\Sigma_{X}}     \nonumber \\   
   \times[(\sigma_{x}^{2} +\gamma_{x})(\sigma_{y}^{2} +\gamma_{y}) - (a+b)^{2}]    
\end{eqnarray*}
Using Eq. (8), this leads to mutual information as

  \begin{eqnarray}
  \mbox{MI}(X,Y)= \frac{-1}{\ln 4} \ln
  \Big(\Big[ 1 - \frac{(a-b)^{2}}{(\sigma_{x}^{2} +\gamma_{x})(\sigma_{y}^{2} +\gamma_{y})}\Big]
  \nonumber  \\
 \times \Big[ 1 - \frac{(a+b)^{2}}{(\sigma_{x}^{2} -\gamma_{x})(\sigma_{y}^{2} -\gamma_{y})}\Big]\Big).
\end{eqnarray}
Eq. (14) suggests that MI decreases monotonically with increasing variances
$\sigma_{x}^{2}$, $\sigma_{y}^{2}$ of $\ln(X)$ and $\ln(Y)$. Moreover, MI has
a minimum as a function of correlations $\gamma_{x}$ between $x_{1}, x_{2}$,
if $2\sigma_{x}^{2}(\sigma_{y}^{2}-\gamma_{y}) \ge (a+b)^{2}$; see Fig. 1
(analogous applies to $\gamma_{y}$). MI also diverges for large intra-correlations
$\gamma_{x}$, $\gamma_{x}$, and for sufficiently strong inter-correlations $a, b$
between $\ln(X)$ and $\ln(Y)$ variables (either positive or negative; Fig. 1).

\begin{figure}[h!]
    \centering
    \includegraphics[scale=0.60]{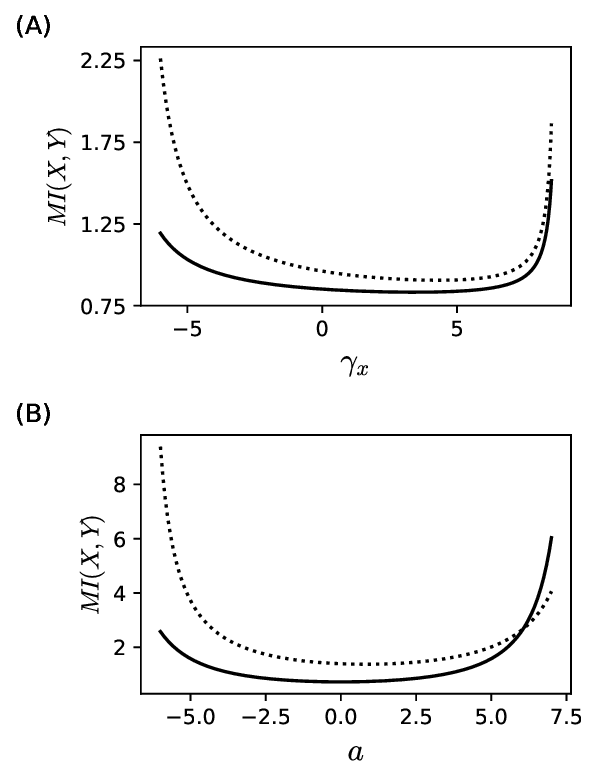}
    \caption{MI in Eq. (20) as a function of $\gamma_x$ and $a$.
     (A) solid line: $\sigma_x^2 = 9$, $\sigma_y^2 = 9$,
      $a = 2.4$, $b = 1$, $\gamma_y = 1$; dotted line: $\sigma_x^2 = 9$,
      $\sigma_y^2 = 9$, $a = 3$, $b = 1.5$, $\gamma_y = 1$.
      (B) solid line: $\sigma_x^2 = 9$, $\sigma_y^2 = 9$, $\gamma_x = 1$,
      $b = 0$, $\gamma_y = 1$; dotted line: $\sigma_x^2 = 9$, $\sigma_y^2 = 16$,
      $\gamma_x = 1$, $b = 8.4$, $\gamma_y = 1$. }
\end{figure}

\subsubsection{Case $n \ge  2$, $k=1$, with tridiagonal matrix for $\Sigma_{X}$.}

In this case we take $\Sigma_{X}$ to be tridiagonal matrix, and choose
$(\Sigma_{X})_{ii}= \sigma_{x}^{2}$, $(\Sigma_{X})_{i,i\pm 1}= \gamma$, with all other
elements of $\Sigma_{X}$ to be zero.
This means that correlated are only those $x_{i}, x_{j}$ that are nearest
neighbors in the ``indices space''. Examples of such short-range correlations are,
e.g., correlations between spins in ferromagnets above a critical temperature,
and density-density correlations in ideal and weakly non-ideal gas (e.g. \cite{sethna}).
Matrix $\Sigma_{Y}$ is just scalar, equal to $\sigma_{y}^{2}$. Covariance
$Cov_{XY}$ is a vector, and we take it $(Cov_{XY})_{i}= a$. This configuration
is depicted by the graph below:

\begin{center}

\begin{tikzpicture}

\draw (0,0) node[draw, circle] (I) {$X_1$}; 
\draw (1.5,0) node[draw, circle] (J) {$X_2$};
\draw (3,0) node[draw, circle] (K) {$X_3$};
\draw (4.5,0) node[draw, circle, scale = 0.75] (A) {$X_{n-1}$};
\draw (6,0) node[draw, circle] (L) {$X_n$};
\node at ($(K)!.5!(A)$) {\ldots};
 
\draw (3,-1.5) node[draw, circle] (M) {$Y$}; 
 
\draw[<->] (I) to (J);
\draw[<->] (J) to (K);
\draw[<->] (L) to (A);

\draw[<->] (L) to (M);
\draw[<->] (K) to (M);
\draw[<->] (J) to (M);
\draw[<->] (I) to (M);
\draw[<->] (A) to (M);

\end{tikzpicture}

\end{center}

The matrix $\Sigma_{X}$ is tridiagonal, and determinants of tridiagonal
matrices are known \cite{hu,molinari}. The determinant
of arbitrary tridiagonal $n\times n$ matrix with all diagonal elements $t$ and all
off-diagonal elements $s$ is given by the function $\theta_{n}(t,s)$, which takes
the form \cite{hu,molinari}

 \begin{eqnarray}
 \theta_{n}(t,s) = 
    \frac{\Big[ \big(t + \sqrt{t^2-4s^2}\big)^{n+1} 
       - \big(t - \sqrt{t^2-4s^2}\big)^{n+1} \Big]}
      {2^{n+1}\sqrt{t^2-4s^2}}.
 \nonumber \\        
    \end{eqnarray}
Hence, we have $\det(\Sigma_{X})=\theta_{n}(\sigma_{x}^{2},\gamma)$.
Mutual information is determined from the first line in Eq. (8), and is given by

  \begin{eqnarray}
  \mbox{MI}(X,Y)= \frac{1}{\ln 4} \ln \left(
  \frac{\sigma_{y}^{2}}
    {\sigma_{y}^{2} - Cov_{XY}^{T}\Sigma_{X}^{-1}Cov_{XY}} \right),
\end{eqnarray}
where $Cov_{XY}^{T}\Sigma_{X}^{-1}Cov_{XY}$ can be found after tedious
calculations as (see the Appendix A)

\begin{widetext}
\begin{eqnarray}
 Cov_{XY}^{T}\Sigma_{X}^{-1}Cov_{XY}=  \frac{na^{2}}{(\sigma_{x}^{2}+2\gamma)}
+ \frac{\gamma a^{2}}{2^{n}(\sigma_{x}^{2}+2\gamma)(\sigma_{x}^{4}-4\gamma^{2})\det(\Sigma_{X})}
\Big[ (-1)^{n+1}2^{n+1}\gamma^{n}(\sigma_{x}^{2}-2\gamma)
\nonumber  \\  
 + (\sigma_{x}^{2}+\sqrt{\sigma_{x}^{4}-4\gamma^{2}})^{n}
  (\sigma_{x}^{2}-2\gamma+\sqrt{\sigma_{x}^{4}-4\gamma^{2}})
  +   (\sigma_{x}^{2}-\sqrt{\sigma_{x}^{4}-4\gamma^{2}})^{n}
 (\sigma_{x}^{2}-2\gamma-\sqrt{\sigma_{x}^{4}-4\gamma^{2}}) \Big].
\end{eqnarray}
\end{widetext}

The above complicated formula, simplifies in two extreme cases.
In the case when $n=2$, we obtain
$Cov_{XY}^{T}\Sigma_{X}^{-1}Cov_{XY}= 2a^{2}/(\sigma_{x}^{2}+\gamma)$, and
hence mutual information takes the simple form 
$\mbox{MI}(X,Y)= \frac{1}{\ln4}
\ln\Big(\frac{\sigma_{y}^{2}}{\sigma_{y}^{2}-2a^{2}/(\sigma_{x}^{2}+\gamma)}\Big)$.
In the case of $n \gg 1$, the first term on the right in Eq. (17) dominates,
and we obtain
$Cov_{XY}^{T}\Sigma_{X}^{-1}Cov_{XY}\approx na^{2}/(\sigma_{x}^{2}+\gamma) + O(1)$,
and MI is given by a similar expression
$\mbox{MI}(X,Y)\approx \frac{1}{\ln4}
\ln\Big(\frac{\sigma_{y}^{2}}{\sigma_{y}^{2}-na^{2}/(\sigma_{x}^{2}+\gamma)}\Big)$.
The latter formula implies that MI diverges for the number of units
$n\approx \sigma_{y}^{2}(\sigma_{x}^{2}+\gamma)/a^{2}$.
In both cases, it is easily seen that MI grows with correlations $a$ between
variables $X$ and $Y$ (both positive and negative), but it decreases monotonically
with increasing the correlations $\gamma$ between $x_{i}$ variables. The latter implies
that negative correlations among $x_{i}$ carry more information than their positive
correlations. This is a different situation than in case number 2, where both strong
positive and negative correlations between $x_{i}$ enhance MI. As expected, MI also
decreases with increasing variances $\sigma_{x}^{2}$ and $\sigma_{y}^{2}$.
General dependence of mutual information on $\gamma$, $\sigma_{x}^{2}$, and $n$,
based on the exact Eqs. (16) and (17) is presented in Fig. 2.

\begin{figure}[h!]
    \centering
    \includegraphics[scale=0.62]{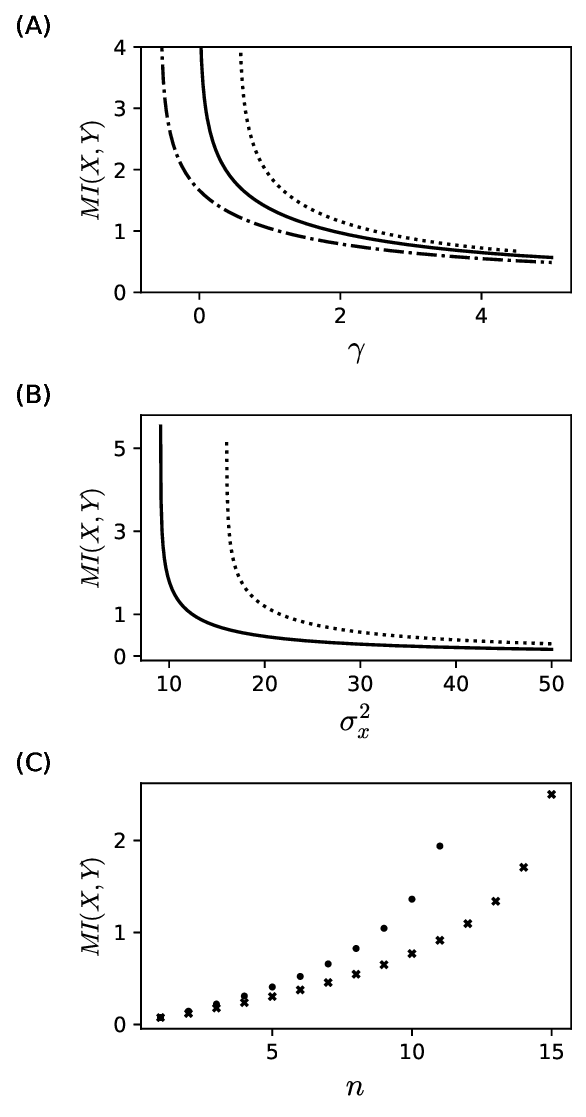}
    \caption{MI in Eq. (22) as a function of $\gamma$, $\sigma_{x}^2$, and
      $n$. (A) solid line: $\sigma_x^2 = 10$,
      $\sigma_y^2 = 1$, $a = 1$, $n = 10$; dotted line: $\sigma_x^2 = 9$,
      $\sigma_y^2 = 9$, $a = 3$, $n = 10$; dash-dotted line: $\sigma_x^2 = 10$,
      $\sigma_y^2 = 10$, $a = 3$, $n = 10$. (B) solid line: $\sigma_y^2 = 1$,
      $a = 1$, $\gamma = \frac{1}{2}$, $n = 10$; dotted line: $\sigma_y^2 = 1$,
      $a = 1.3$, $\gamma = \frac{1}{2}$, $n = 10$. (C) points: $\sigma_x^2 = 10$,
      $\sigma_y^2 = 1$, $a = 1$, $\gamma = 1$; x: $\sigma_x^2 = 10$, $\sigma_y^2 = 1$,
      $a = 1$, $\gamma = 3$.}
\end{figure}

\subsubsection{Case $n \ge  2$, $k=1$, with Kac-Murdock-Szeg{\"o} matrix for $\Sigma_{X}$.}

The only difference with the point 3 above is that now the matrix $\Sigma_{X}$ has all
nonzero elements, which are given by $(\Sigma_{X})_{ij}= \sigma_{x}^{2}\gamma^{|i-j|}$,
where $|\gamma| \le 1$ \cite{kac}. This means that all $x_{i}$ are mutually correlated,
but distant ones (with remote indices) are gradually less correlated (see below).

\begin{center}

\begin{tikzpicture}

\draw (0,0) node[draw, circle] (I) {$X_1$}; 
\draw (1.5,0) node[draw, circle] (J) {$X_2$};
\draw (3,0) node[draw, circle] (K) {$X_3$};
\draw (4.5,0) node[draw, circle, scale = 0.75] (A) {$X_{n-1}$};
\draw (6,0) node[draw, circle] (L) {$X_n$};
\node at ($(K)!.5!(A)$) {\ldots};
 
\draw (3,-1.5) node[draw, circle] (M) {$Y$}; 
 
\draw[<->] (I) to (J);
\draw[<->] (J) to (K);
\draw[<->] (L) to (A);

\draw[<->] (L) to (M);
\draw[<->] (K) to (M);
\draw[<->] (J) to (M);
\draw[<->] (I) to (M);
\draw[<->] (A) to (M);

\draw[<->] (I) edge [bend left] (K);
\draw[<->] (I) edge [bend left] (A);
\draw[<->] (I) edge [bend left] (L);

\draw[<->, dashed] (J) edge [bend left] (K);
\draw[<->, dashed] (J) edge [bend left] (A);
\draw[<->, dashed] (J) edge [bend left] (L);

\draw[<->, dashed] (K) edge [bend right] (A);
\draw[<->, dashed] (K) edge [bend right] (L);

\end{tikzpicture}

\end{center}

The inverse of the matrix $\Sigma_{X}$ is \cite{kac,horn}

\begin{eqnarray}
  (\Sigma_{X}^{-1})_{ij}= \frac
  {\delta_{ij}[1+\gamma^{2}(1-\delta_{i1}-\delta_{in})]
  -\gamma(\delta_{i,j-1}+\delta_{i,j+1})}
  {\sigma_{x}^{2}(1-\gamma^{2})},   \nonumber \\
\end{eqnarray}
which allows us to find
\begin{eqnarray}
  Cov_{XY}^{T}\Sigma_{X}^{-1}Cov_{XY}=
  \frac{a^{2}[n(1-\gamma)+2\gamma]}{\sigma_{x}^{2}(1+\gamma)}. 
\end{eqnarray}
Consequently, mutual information in this case is given by a simple
explicit formula:

\begin{eqnarray}
 \mbox{MI}(X,Y)= \frac{1}{\ln 4} \ln \left(
  \frac{\sigma_{y}^{2}}
    {\sigma_{y}^{2} - \frac{a^{2}[n(1-\gamma)+2\gamma]}{\sigma_{x}^{2}(1+\gamma)} }
     \right).
\end{eqnarray}
It is easily seen that MI grows monotonically with the number of units $n$
and diverges for
$n= \frac{[\sigma_{x}^{2}\sigma_{y}^{2}(1+\gamma) -2\gamma a^{2}]}{(1-\gamma)a^{2}}$.
The dependence of MI on other parameters is similar to the case number 3 above.
In particular, MI decreases monotonically with increasing correlations $\gamma$
among $x_{i}$ variables from negative ($\gamma < 0$) to positive ($\gamma > 0$).

\subsubsection{Case $n \ge 2$, $k \ge 2$,  with non-symmetric and nondiagonal $Cov_{XY}$.}

Let us consider $n\times n$ matrix $\Sigma_{X}$ and $k\times k$ matrix $\Sigma_{Y}$ to
be both tridiagonal with elements 
$(\Sigma_{X})_{ij}= \sigma_{x}^{2}\delta_{ij} + \gamma_{x}(\delta_{i,j-1}+\delta_{i,j+1})$,
and $\Sigma_{Y}= \sigma_{y}^{2}\delta_{ij} + \gamma_{y}(\delta_{i,j-1}+\delta_{i,j+1})$.
The covariance $Cov_{XY}$ is $n\times k$ sparse matrix taken with one nonzero element
$(Cov_{XY})_{n1}=a$ and the rest elements are 0 (see the graph below).

\begin{figure}[h]
\centering
\begin{tikzpicture}

\draw (0,0) node[draw, circle] (I) {$X_1$}; 
\draw (1.5,0) node[draw, circle] (J) {$X_2$};
\draw (3,0) node[draw, circle] (K) {$X_3$};
\draw (4.5,0) node[draw, circle, scale=0.75] (L) {$X_{n-1}$};
\draw (6,0) node[draw, circle] (Z) {$X_n$};
\node at ($(K)!.5!(L)$) {\ldots};
 
\draw (0,-2.5) node[draw, circle] (M) {$Y_1$}; 
\draw (1.5,-2.5) node[draw, circle] (N) {$Y_2$};
\draw (3,-2.5) node[draw, circle] (O) {$Y_3$};
\draw (4.5,-2.5) node[draw, circle, scale=0.75] (P) {$Y_{k-1}$};
\draw (6,-2.5) node[draw, circle] (V) {$Y_k$};
\node at ($(O)!.5!(P)$) {\ldots};
 
 \draw[<->] (I) to (J);
 \draw[<->] (J) to (K);
 \draw[<->] (L) to (Z);

 \draw[<->] (M) to (N);
 \draw[<->] (N) to (O);
 \draw[<->] (P) to (V);

\draw[<->] (Z) to (M);

\end{tikzpicture}
\end{figure}

This case is relevant for two groups of cascade networks
of interacting elements.

We use the last line in Eq. (8) for MI. In this configuration, it is easy to show
that $(\Sigma_{X}^{-1}Cov_{XY}\Sigma_{Y}^{-1}Cov_{XY}^{T})_{ij}=
a^{2}(\Sigma_{X}^{-1})_{in}(\Sigma_{Y}^{-1})_{11}$ for $j=n$ and 0 for $1 \le j \le n-1$,
for every $i$. This leads to the simple form for the determinant
$\det(I - \Sigma_{X}^{-1}Cov_{XY}\Sigma_{Y}^{-1}Cov_{XY}^{T})=
1 -a^{2}(\Sigma_{X}^{-1})_{nn}(\Sigma_{Y}^{-1})_{11}$, and consequently for MI

\begin{eqnarray}
 \mbox{MI}(X,Y)= \frac{1}{\ln 4} \ln \Big(
  \frac{1}{1 -a^{2}(\Sigma_{X}^{-1})_{nn}(\Sigma_{Y}^{-1})_{11}}
     \Big).
\end{eqnarray}
The matrix elements $(\Sigma_{X}^{-1})_{nn}$ and $(\Sigma_{Y}^{-1})_{11}$
are given by the ratios of the function $\theta_{n}$ defined in Eq. (15), i.e.,
$(\Sigma_{X}^{-1})_{nn}= \theta_{n-1}(\sigma^{2}_{x},\gamma_{x})/\theta_{n}(\sigma^{2}_{x},\gamma_{x})$,
and
$(\Sigma_{Y}^{-1})_{11}= \theta_{k-1}(\sigma^{2}_{y},\gamma_{y})/\theta_{k}(\sigma^{2}_{y},\gamma_{y})$;
see Appendix A.

In the limits $n \gg 1$, $k \gg 1$, and for $\sigma_{x}^{2} > 2\gamma_{x}$
and $\sigma_{y}^{2} > 2\gamma_{y}$, we obtain an approximated MI as

\begin{eqnarray*}
 \mbox{MI}(X,Y)_{n\gg 1} \approx \frac{-1}{\ln 4} \ln \Big(
  1 -\frac{4a^{2}}
  {\big(\sigma_{x}^{2}+\sqrt{\sigma_{x}^{4}-4\gamma_{x}^{2}}\big)}
  \\
  \times \big(\sigma_{y}^{2}+\sqrt{\sigma_{y}^{4}-4\gamma_{y}^{2}}\big)^{-1} 
     \Big).
\end{eqnarray*}
This equation shows that large variances of both $\sigma_{x}^{2}$ and
$\sigma_{y}^{2}$ have detrimental effect on MI. On the other hand,
increasing covariances among $\ln(X)$ ($|\gamma_{x}|$), and among $\ln(Y)$
($|\gamma_{y}|$) causes increase in MI.

\subsubsection{Case $n = k \ge 2$ with symmetric and diagonal $Cov_{XY}$.}

In this case, we take $\Sigma_{X}$ to be tridiagonal with elements 
$(\Sigma_{X})_{ij}= \sigma_{x}^{2}\delta_{ij} + \gamma_{x}(\delta_{i,j-1}+\delta_{i,j+1})$,
and $\Sigma_{Y}$ to be diagonal with elements 
$(\Sigma_{Y})_{ij}= \sigma_{y}^{2}\delta_{ij}$. Additionally, we choose the
covariance matrix $Cov_{XY}$ to be diagonal, $(Cov_{XY})_{ij}= a\delta_{ij}$.
This situation corresponds to the system depicted below

\begin{center}

\begin{tikzpicture}

\draw (0,0) node[draw, circle] (A) {$X_1$}; 
\draw (1.5,0) node[draw, circle] (B) {$X_2$};
\draw (3,0) node[draw, circle] (C) {$X_3$};
\draw (4.5,0) node[draw, circle, scale = 0.75] (D) {$X_{n-1}$};
\draw (6,0) node[draw, circle] (Z) {$X_n$};
\node at ($(C)!.5!(D)$) {\ldots};
 
\draw (0,-1.5) node[draw, circle] (E) {$Y_1$}; 
\draw (1.5,-1.5) node[draw, circle] (F) {$Y_2$};
\draw (3,-1.5) node[draw, circle] (G) {$Y_3$};
\draw (4.5,-1.5) node[draw, circle, scale=0.75] (X) {$Y_{n-1}$};
\draw (6,-1.5) node[draw, circle] (H) {$Y_n$};
\node at ($(G)!.5!(X)$) {\ldots};
 
\draw[<->] (A) to (B);
\draw[<->] (B) to (C);
\draw[<->] (D) to (Z);

\draw[<->] (A) to (E);
\draw[<->] (B) to (F);
\draw[<->] (C) to (G);
\draw[<->] (D) to (X);
\draw[<->] (Z) to (H);

\end{tikzpicture}

\end{center}

Here, there are correlations between nearest neighbors of variables $x_{i}$, but there
are no correlations between variables $y_{i}$. Additionally, $x_{i}$ correlates directly
only with $y_{i}$.

Elements of the matrix of interest   $\Sigma_{X}-Cov_{XY}\Sigma_{Y}^{-1}Cov_{XY}^{T}$
appearing in Eq. (8) for mutual information are
$(\Sigma_{X}-Cov_{XY}\Sigma_{Y}^{-1}Cov_{XY}^{T})_{ij}=
(\sigma_{x}^{2}-\frac{a^{2}}{\sigma_{y}^{2}})\delta_{ij}
+ \gamma_{x}(\delta_{i,j-1} + \delta_{i,j+1})$. Thus it is also the tridiagonal
matrix, similar as $\Sigma_{X}$. Therefore, the determinants of both $\Sigma_{X}$
and $\Sigma_{X}-Cov_{XY}\Sigma_{Y}^{-1}Cov_{XY}^{T}$ are given by Eq. (15),
i.e., $\det(\Sigma_{X})= \theta_{n}(\sigma^{2}_{x},\gamma_{x})$ and
$\det{(\Sigma_{X}-Cov_{XY}\Sigma_{Y}^{-1}Cov_{XY}^{T})}=
\theta_{n}(\sigma_{x}^{2}-\frac{a^{2}}{\sigma_{y}^{2}},\gamma_{x})$.

The corresponding mutual information is given by the first line in Eq. (8) as

\begin{eqnarray}
 \mbox{MI}(X,Y)= \frac{1}{\ln 4} \ln \left(
  \frac{\theta_{n}(\sigma^{2}_{x},\gamma_{x})}
     {\theta_{n}(\sigma_{x}^{2}-\frac{a^{2}}{\sigma_{y}^{2}},\gamma_{x}) }
     \right).
\end{eqnarray}
The dependence of MI on $\gamma_{x}$, $a$, and $n$ is shown in Fig. 3.
In a special case $n=k=2$, we recover MI in the case
number 2 (see Eq. (14)), but with $b= \gamma_{y}= 0$.

\begin{figure}[h!]
    \centering
    \includegraphics[scale=0.62]{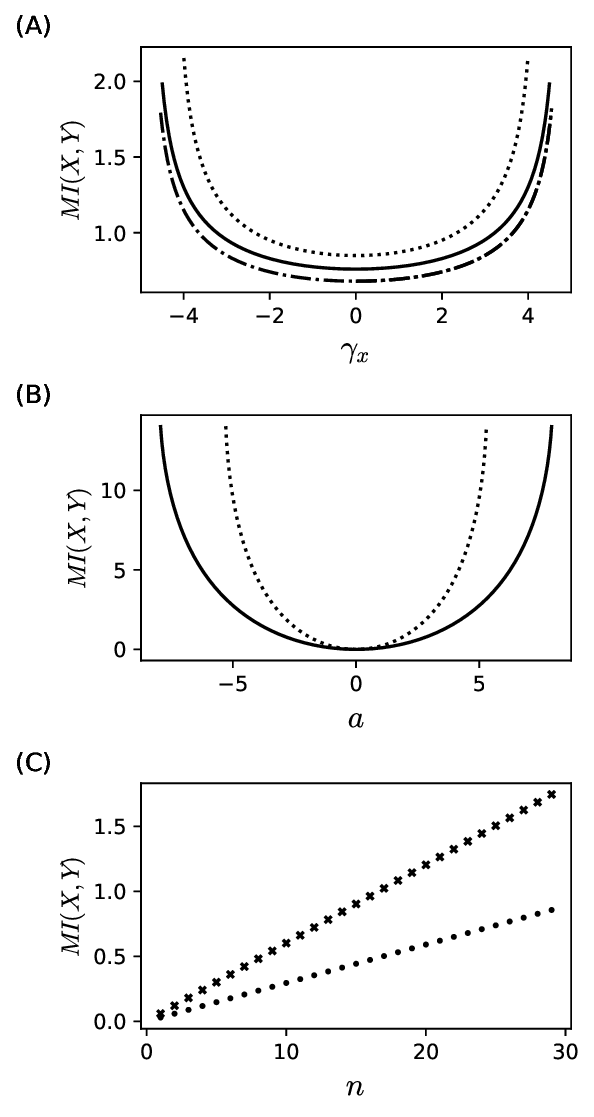}
    \caption{MI in Eq. (28) as a function of $\gamma_x$, $a$,
      and $n$.
      (A) solid line: $\sigma_x^2 = 10$, $\sigma_y^2 = 1$, $a = 1$, $n = 10$;
      dotted line: $\sigma_x^2 = 9$, $\sigma_y^2 = 9$, $a = 3$, $n = 10$;
      dash-dotted line: $\sigma_x^2 = 10$, $\sigma_y^2 = 10$, $a = 3$, $n = 10$.
      (B) solid line: $\sigma_x^2 = 9$, $\sigma_y^2 = 9$, $\gamma_x = 1$, $n = 10$;
      dotted line: $\sigma_x^2 = 9$, $\sigma_y^2 = 4$, $\gamma_x = 1$, $n = 10$.
      (C) points: $\sigma_x^2 = 25$, $\sigma_y^2 = 1$, $a = 1$, $\gamma_x = 1$;
      x: $\sigma_x^2 = 50$, $\sigma_y^2 = 1$, $a = 2$, $\gamma_x = 2$.}
    \label{fig:enter-label}
\end{figure}

\section{\label{sec:level3} Information transfer for neurons}

Here, we apply the exact formula for MI in Eq. (8) to neuroscience, i.e.,
we derive mutual information relevant for information transfer in neural networks
of a mammalian brain. Specifically, we find MI between activities of $n$ weakly
correlated presynaptic neurons with firing rates $\vec{f}=(f_{1},f_{2},...,f_{n})$
and activity of a postsynaptic neuron with the firing rate $Y$. This situation can be
depicted by the analogical graph as in the case no. 4 considered above. Below we perform
computations and make comparison for two cases: one assuming that presynaptic
firing rates $\vec{f}$ are lognormally distributed, and second that they are
normally distributed. We note that although the mutual information between pre-
and post-synaptic neural activities has been found many times in the past
\cite{rieke,paninski,borst,brunel,panzeri,levy}, none of those approaches used
lognormally distributed variables (either Gaussian or a mixture of specific, like
Poisson, and non-specific discrete distributions were used).

The simplest model relating post- and presynaptic firing rates $Y$ and $\vec{f}$ is
\cite{hertz,dayan}

\begin{eqnarray}
Y= \sum_{i=1}^{n} w_{i}f_{i},
\end{eqnarray}
where $\vec{w}=(w_{1},w_{2},...,w_{n})$ are synaptic weights, and we assume that all
of them are positive (all excitatory presynaptic neurons). For cortical neurons the
number of synaptic contacts $n$ is very large, typically in the range
$n\sim 10^{3}-10^{4}$ \cite{braitenberg}.
The dependence of $Y$ on $\vec{f}$ could in principle be nonlinear.
We assume linearity, since it was shown that for a biophysically motivated nonlinear
class of neurons (so-called class 2), the dependence of $Y$ on $\vec{f}$ can become
linear by incorporating adaptation in neural firing rates, which is often observed in
brain networks \cite{ermentrout1998,ermentrout2010,karbowski}.

\subsection{\label{sec:levell3} MI for neural activities as lognormal variables.}

Our major assumption in this section is that both firing rates $\vec{f}$ and
synaptic weights $\vec{w}$ have lognormal distributions, which is compatible with
experimental data for neurons and synapses in the mammalian cerebral cortex
\cite{song,hromadka,buzsaki}. Additionally, we assume that all $f_{i}$ (and
corresponding $w_{i}$) have the same parameters characterizing the distribution
(uniformity assumption).
Thus the postsynaptic firing rate $Y$ in Eq. (23) is a sum of lognormal random
variables identically distributed, since each product $w_{i}f_{i}$ is
lognormally distributed. It should be clearly said that $Y$ has an unknown
exact form of probability density. However, it has been numerically verified by
others \cite{abu-dayya,mehta} that the sum of (uncorrelated and correlated)
lognormal random variables can be approximated by lognormal distribution for
large but finite $n$. Moreover and more importantly, it has been proven that in
the limiting case $n\mapsto \infty$ the sum of positively correlated lognormals also
has a lognormal probability density \cite{szyszkowicz,beaulieu}.
(Standard Central Limit Theorem does not apply here because of the correlations
between the summands with heavy tails.)
For these reasons, in this section, we assume that postsynaptic neural activity
$Y$ has a lognormal distribution.

The goal is to find mutual information $\mbox{MI}(\vec{f},Y)$ between vectors $\vec{f}$
and $Y$. We assume that activities of presynaptic neurons (i.e., $\ln(f_{i})$)
are weakly correlated, and take for their covariance matrix
$\Sigma_{f}$ the form given by Kac-Murdock-Szeg{\"o} matrix,
i.e., $(\Sigma_{f})_{ij}= \sigma_{f}^{2}\gamma^{|i-j|}$,
with $|\gamma| \ll  1$, where $\sigma_{f}^{2}$ is the variance for all $\ln(f_{i})$.
The corresponding variance for all $\ln(w_{i})$ is denoted as $\sigma_{w}^{2}$,
and the variance for postsynaptic activity $\ln(Y)$ as $\sigma_{Y}^{2}$.
The latter depends on the parameters characterizing distributions
of $\vec{f}$ and $\vec{w}$ (see below).
Additionally, we assume that synaptic weights $w_{i}$ are not correlated
between themselves, and that vectors $\vec{f}$ and $\vec{w}$ are not correlated either.
These two assumptions are consistent with empirical observations, but are not essential
for the derivation (in principle they could be included). They make the final
formula look simpler. Using Eq. (2), we can write expressions for various moments
of $f_{i}$ and $w_{i}$, which are used later:

\begin{eqnarray}
  \langle f_{i}\rangle \equiv \langle f\rangle
  = e^{\mu_{f} + \frac{1}{2}\sigma_{f}^{2}},   \nonumber \\
  \langle w_{i}\rangle \equiv \langle w\rangle
  = e^{\mu_{w} + \frac{1}{2}\sigma_{w}^{2}},   \nonumber \\
  \langle f_{i}^{2}\rangle \equiv \langle f^{2}\rangle 
  = e^{2\mu_{f} + 2\sigma_{f}^{2}},   \nonumber \\
  \langle w_{i}^{2}\rangle \equiv  \langle w^{2}\rangle
  = e^{2\mu_{w} + 2\sigma_{w}^{2}},   \nonumber \\
 \langle f_{i}f_{j}\rangle = e^{2\mu_{f} + \sigma_{f}^{2}(1+c_{ij})},   
\end{eqnarray}
where we used the uniformity of the moments, and consequently the uniformity of
the parameters
$\mu_{f}= \langle\ln(f_{i})\rangle$, $\mu_{w}= \langle\ln(w_{i})\rangle$.
Additionally, $c_{ij}$ is the correlation coefficient between
$\ln(f_{i})$ and $\ln(f_{j})$, i.e.,
$c_{ij}\sigma_{f}^{2} =\langle(\ln(f_{i})-\mu_{f})(\ln(f_{j})-\mu_{f})\rangle=
(\Sigma_{f})_{ij}$. The latter means that $c_{ij}= \gamma^{|i-j|}$.

To determine MI we need the variance of $\ln(Y)$ and covariance matrix between
$\ln(Y)$ and $\ln(f_{i})$. This is accomplished by finding the first two moments
of $Y$ in terms of the parameters characterizing $\vec{f}$ and $\vec{w}$, and
match them to the first two moments of assumed lognormal form of $Y$
(so-called Wilkinson method; \cite{abu-dayya,mehta}). We have

\begin{eqnarray*}
 \langle Y\rangle = \sum_{i=1}^{n}  \langle w_{i}\rangle \langle f_{i}\rangle,
\end{eqnarray*}
and
\begin{eqnarray*}
  \langle Y^{2}\rangle = \sum_{i=1}^{n} \langle w_{i}^{2}\rangle \langle f_{i}^{2}\rangle
   + 2 \sum_{i=1}^{n-1}\sum_{j=i+1}^{n} \langle w_{i}\rangle \langle w_{j}\rangle
   \langle f_{i}f_{j}\rangle,
\end{eqnarray*}
which after using Eqs. (24) yields

\begin{eqnarray}
  \langle Y\rangle = ne^{\mu_{f}+\mu_{w}} e^{(\sigma_{f}^{2}+\sigma_{w}^{2})/2},   
  \nonumber \\
  \langle Y^{2}\rangle = e^{2(\mu_{f}+\mu_{w})} e^{\sigma_{f}^{2}+\sigma_{w}^{2}}
   \nonumber \\  
 \times\Big( ne^{\sigma_{f}^{2}+\sigma_{w}^{2}} 
   + 2 \sum_{i=1}^{n-1}\sum_{j=i+1}^{n} e^{\sigma_{f}^{2}c_{ij}} \Big).
\end{eqnarray}
On the other hand, for lognormally distributed $Y$, we have
$\langle Y\rangle = e^{\mu_{Y} + \frac{1}{2}\sigma_{Y}^{2}}$, and  
$\langle Y^{2}\rangle = e^{2\mu_{Y} + 2\sigma_{Y}^{2}}$, where $\mu_{Y}$ and
$\sigma_{Y}^{2}$ are the mean and the variance of $\ln(Y)$. 
This allows us to find $\sigma_{Y}^{2}$ in terms of the parameters for $\vec{f}$
and $\vec{w}$ as

\begin{eqnarray}
\sigma_{Y}^{2} = \ln\Big(\frac{1}{n} e^{(\sigma_{f}^{2}+\sigma_{w}^{2})}
   + \frac{2}{n^{2}} \sum_{i=1}^{n-1}\sum_{j=i+1}^{n} e^{\sigma_{f}^{2}c_{ij}} \Big).
\end{eqnarray}

To find the covariance $Cov_{Y\vec{f}}\equiv Cov(\ln(Y),\ln(\vec{f}))$
between $\ln(Y)$ and $\ln(f_{i})$, we first determine
the covariance $Cov(Y,f_{i})$ between $Y$ and $f_{i}$. We have 
$Cov(Y,f_{i})= \langle Yf_{i}\rangle  - \langle Y\rangle\langle f_{i}\rangle
= \langle w_{i}\rangle\langle f_{i}^{2}\rangle
- \langle Y\rangle\langle f_{i}\rangle
+ \sum_{j\neq i}^{n} \langle w_{j}\rangle\langle f_{i}f_{j}\rangle$.
All these moments are given above, and we obtain

\begin{eqnarray}
 Cov(Y,f_{i})= 
 e^{m_{w}+\sigma_{w}^{2}/2}e^{2(m_{f}+\sigma_{f}^{2})}  \nonumber \\
\times \Big( 1 - e^{-\sigma_{f}^{2}}\big[n - \sum_{j\neq i}^{n} e^{\sigma_{f}^{2}c_{ij}}\big] \Big),
\end{eqnarray}
which after using Eq. (2) yields the covariance vector of interest

\begin{eqnarray}
(Cov_{Y\vec{f}})_{i}=
  \ln\Big(\frac{1}{n} \big[e^{\sigma_{f}^{2}}
   + \sum_{j\neq i}^{n} e^{\sigma_{f}^{2}c_{ij}}\big] \Big).
\end{eqnarray}
The general formulas in Eqs. (26) and (28) can be expanded for weak
presynaptic neural correlations,
i.e., $e^{\sigma_{f}^{2}c_{ij}}\approx 1+ \sigma_{f}^{2}c_{ij}$
for $|\gamma| \ll 1$, with the help of
$\sum_{j\neq i} c_{ij}= \gamma
\big[\delta_{i1}+\delta_{in}+2(1-\delta_{i1})(1-\delta_{in})\big] + O(\gamma^{2})$.
This leads to

\begin{eqnarray}
\sigma_{Y}^{2} = \ln\Big(1 + \frac{1}{n}\big[ e^{(\sigma_{f}^{2}+\sigma_{w}^{2})} -1\big]\Big)
 \nonumber \\
+ \frac{2\gamma\sigma_{f}^{2}(1-\frac{1}{n})^{2}}
{n+ [e^{(\sigma_{f}^{2}+\sigma_{w}^{2})} -1]} + O(\gamma^{2}),
\end{eqnarray}
and

\begin{eqnarray}
(Cov_{Y\vec{f}})_{i}=
\ln\Big(1 + \frac{1}{n}\big[ e^{\sigma_{f}^{2}} -1\big]\Big)
 \nonumber \\
+ \frac{\gamma\sigma_{f}^{2}\big[\delta_{i1}+\delta_{in}+2(1-\delta_{i1})(1-\delta_{in})\big]}
{n + [e^{\sigma_{f}^{2}} -1]} + O(\gamma^{2}).
\end{eqnarray}
Note that the covariance vector $Cov_{Y\vec{f}}$ is finite even when there are no
correlations between presynaptic neural activities ($\gamma\mapsto 0$),
however in the limit $n\mapsto\infty$ it vanishes.

The last step is to find the value of $Cov_{Y\vec{f}}^{T}\Sigma_{f}^{-1}Cov_{Y\vec{f}}$,
using approximation to Eq. (18) for small $\gamma$, i.e., 
$(\Sigma_{f}^{-1})_{ij}= \frac{1}{\sigma_{f}^{2}}\big[\delta_{ij}-
  \gamma(\delta_{i,j-1} + \delta_{i,j+1})\big] + O(\gamma^{2})$.
The result is

\begin{eqnarray}
Cov_{Y\vec{f}}^{T}\Sigma_{f}^{-1}Cov_{Y\vec{f}}=
 \frac{n(1-2\gamma)}{\sigma_{f}^{2}}
  \ln^{2}\Big(1 + \frac{1}{n}\big[e^{\sigma_{f}^{2}} -1\big]\Big)
\nonumber  \\
+ \frac{4(n-1)\gamma}{(e^{\sigma_{f}^{2}} + n-1)}
\ln\Big(1 + \frac{1}{n}\big[e^{\sigma_{f}^{2}} -1\big]\Big)
+ O(\gamma^{2}).
\nonumber  \\
\end{eqnarray}
This term also vanishes in the limit of very large number of presynaptic
neurons, $n\mapsto \infty$.

Mutual information between pre- and post-synaptic neuronal activities
is given by the first line of Eq. (8), with $\det(\Sigma_{Y})= \sigma_{Y}^{2}$,
and
$\det\Big(\Sigma_{Y}-Cov_{Y\vec{f}}^{T}\Sigma_{f}^{-1}Cov_{Y\vec{f}}\Big)=
\sigma_{Y}^{2}- Cov_{Y\vec{f}}^{T}\Sigma_{f}^{-1}Cov_{Y\vec{f}}$,
which are given by Eq. (29) and (31). MI takes the form

\begin{widetext}
\begin{eqnarray}
 \mbox{MI}(\vec{f},Y)_{ln}= \frac{1}{\ln4} \ln\Big(
\frac{ \ln\big(1 + \frac{1}{n}[e^{(\sigma_{f}^{2}+\sigma_{w}^{2})} -1]\big) }
     {\big[\ln\big(1 + \frac{1}{n}[e^{(\sigma_{f}^{2}+\sigma_{w}^{2})} -1]\big)
  -  \frac{n}{\sigma_{f}^{2}}
  \ln^{2}\big(1 + \frac{1}{n}\big[e^{\sigma_{f}^{2}} -1\big]\big) \big]}
\Big[ 1 + \gamma g(n,\sigma_{f}^{2},\sigma_{w}^{2}) + O(\gamma^{2}) \Big]
     \Big),
\end{eqnarray}
where the function $g(n,\sigma_{f}^{2},\sigma_{w}^{2})$ is defined as 

\begin{eqnarray}
 g(n,\sigma_{f}^{2},\sigma_{w}^{2})= 
 \frac{ 2n \ln^{2}\big(1 + \frac{1}{n}[e^{\sigma_{f}^{2}}-1]\big)
   \Big[ \Big( \frac{2(1-1/n)}{(e^{\sigma_{f}^{2}} + n-1)
   \ln\big(1 + \frac{1}{n}[e^{\sigma_{f}^{2}}-1]\big)}
     - \frac{1}{\sigma_{f}^{2}} \Big)
 \ln\big(1 + \frac{1}{n}[e^{(\sigma_{f}^{2}+\sigma_{w}^{2})} -1]\big)
 -  \frac{(1-1/n)}{\big(e^{\sigma_{f}^{2}+\sigma_{w}^{2}} +n-1\big)} \Big] }
     {\ln\big(1 + \frac{1}{n}[e^{(\sigma_{f}^{2}+\sigma_{w}^{2})} -1]\big)
\Big[ \ln\big(1 + \frac{1}{n}[e^{(\sigma_{f}^{2}+\sigma_{w}^{2})} -1]\big) 
  - \frac{n}{\sigma_{f}^{2}} \ln^{2}\big(1 + \frac{1}{n}\big[e^{\sigma_{f}^{2}} -1\big]\big) \Big] },
 \nonumber \\
\end{eqnarray}
\end{widetext}
with $\sigma_{f}^{2}= \ln(1+Var(f)/\langle f\rangle^{2})$ and
$\sigma_{w}^{2}= \ln(1+Var(w)/\langle w\rangle^{2})$, where  
$Var(f)= \langle f^{2}\rangle - \langle f\rangle^{2}$, and
$Var(w)= \langle w^{2}\rangle - \langle w\rangle^{2}$ 
(see Eq. (24)). Eqs. (32) and (33) constitute the major result of this paper.
These two equations with quite complex formulas give us MI$_{ln}$ for any number
of presynaptic neurons $n$, as well as for arbitrary levels of variability
in presynaptic firing rates $\sigma_{f}$ and synaptic weights $\sigma_{w}$.
Note that MI$_{ln}$ does not depend at all on $\mu_{f}$ and $\mu_{w}$.
For very noisy synapses, i.e., for $\sigma_{w}\mapsto \infty$, both factors under
the logarithm in Eq. (32) tend to 1 (function $g\mapsto 0$), and hence mutual
information $\mbox{MI}_{ln}\mapsto 0$. The function $g$ in Eq. (33) is generally
positive (unless presynaptic activity has very high variability, $\sigma_{f}^{2} \gg 1$,
but then MI already approaches infinity). Specifically, for $\sigma_{f}^{2} \ll 1$,
and arbitrary $\sigma_{w}^{2}$ we get a relatively simple form for this function,
$g\approx 2\sigma_{f}^{2}/\big[n\ln(1+[e^{\sigma_{w}^{2}}-1]/n)\big]$, which
implies that positive correlations between presynaptic neurons ($\gamma > 0$)
increase MI, while their negative correlations ($\gamma < 0$) are detrimental
for MI.

Of particular interest is the expression for MI when neuronal activities
and synaptic weights have low variability, such that both
$\sigma_{f}^{2} \ll 1$ and $\sigma_{w}^{2} \ll 1$.
In this case, Eqs. (32) and (33) simplify significantly, and we obtain
the simple expression for MI

\begin{eqnarray}
\mbox{MI}(\vec{f},Y)_{ln} \approx
\frac{\ln\Big( \Big[ 1 + \frac{\sigma_{f}^{2}}{\sigma_{w}^{2}}\Big]
\Big[1 + \frac{2\gamma\sigma_{f}^{2}}{(\sigma_{w}^{2}+\sigma_{f}^{2})} + O(\gamma^{2}) \Big]
\Big)}{\ln4}.
\nonumber \\
\end{eqnarray}
Note that MI in this limit is independent of the number of presynaptic neurons $n$.
Moreover, Eq. (34) implies that MI grows with increasing the variance of presynaptic
neurons $\sigma_{f}^{2}$, and it decays with increasing synaptic weight variance
$\sigma_{w}^{2}$. This suggests that the ratio $\sigma_{f}^{2}/\sigma_{w}^{2}$ can
be interpreted as the signal to noise ratio, with the variability in synaptic
weights serving as the noise.

Eq. (34) can be rewritten in terms of means and variances of presynaptic firings
$\langle f\rangle$, $Var(f)$, and of synaptic weights
$\langle w\rangle$, $Var(w)$, using Eq. (24).
For $\sigma_{f}^{2}  \ll 1$ and $\sigma_{w}^{2}  \ll 1$,
we have $\sigma_{f}^{2}\approx Var(f)/\langle f\rangle^{2}$, and
$\sigma_{w}^{2}\approx Var(w)/\langle w\rangle^{2}$, which leads
to

\begin{eqnarray}
\mbox{MI}(\vec{f},Y)_{ln} \approx
\frac{1}{\ln4}\ln\Big( \Big[ 1 + \frac{\langle w\rangle^{2} Var(f)}
  {\langle f\rangle^{2} Var(w)} \Big]
  \nonumber \\
\times \Big[ 1 + \frac{2\gamma\langle w\rangle^{2} Var(f)}
  {[\langle f\rangle^{2} Var(w) + \langle w\rangle^{2} Var(f)]} + O(\gamma^{2}) \Big] \Big).
\end{eqnarray}
From this formula it is clear that in the limit of low neuronal and
synaptic variabilities, MI scales with their relative ratio.

\subsection{\label{sec:levell3} Comparison between neuronal MI with lognormal and normal
  neural activities.}

In this section we take a more traditional viewpoint and derive neuronal MI treating
neural activities as Gaussian variables. The resulting expression for Gaussian MI is then
compared with lognormal MI obtained in Eq. (32).

It should be said from the outset that treating positive firing rates as normal
variables is a little unrealistic, since there is always some nonzero probability that the
rates can become negative. Obviously, that likelihood is extremely small if neural activity
variances are much smaller than their mean levels.

We assume explicitly that firing rates $\vec{f}$ in Eq. (23) have normal distribution
with uniform mean $\langle f\rangle$ and variance $Var(f)$, and they are weakly
correlated. No specific distribution for synaptic weights $\vec{w}$ is assumed,
except that it has uniform mean $\langle w\rangle$ and finite but not too large
variance $Var(w)$ (it could be Gaussian too, but it is not necessary for the arguments
below). We also take the limit $n\mapsto \infty$, implying very large
number of presynaptic neurons. That limit allows us to use the central limit
theorem in Eq. (23) and claim that the postsynaptic firing rate $Y$ has a normal
distribution. The Central Limit Theorem is permissible here because, in contrast
to the lognormal case, the summands in Eq. (23) have short tails \cite{vankampen}.

The goal is to find mutual information between Gaussian $\vec{f}$ and $Y$, denoted
as $\mbox{MI}(\vec{f},Y)_{g}$, which is given by the first line of Eq. (8), where
$X \mapsto \vec{f}$. This means that now
$\Sigma_{Y}= \langle Y^{2}\rangle - \langle Y\rangle^{2}$,
covariance vector between $Y$ and $\vec{f}$ is
$Cov(Y,f_{i})= \langle Yf_{i}\rangle - \langle Y\rangle\langle f_{i}\rangle$,
and covariance matrix $\Sigma_{f}$ is between $f_{i}$ and $f_{j}$, i.e.,
$(\Sigma_{f})_{ij}= \langle f_{i}f_{j}\rangle - \langle f_{i}\rangle\langle f_{j}\rangle$.
Again, for the latter matrix we take the Kac-Murdock-Szeg{\"o} form, though with
different coefficients,  $(\Sigma_{f})_{ij}= c_{0}\kappa^{|i-j|}$, where
the variance in presynaptic activities $c_{0}= \langle f^{2}\rangle - \langle f\rangle^{2}$,
and $|\kappa| \ll 1$. The inverse of $\Sigma_{f}$ is
$(\Sigma_{f}^{-1})_{ij}= [\delta_{ij}-\kappa(\delta_{i,j-1}+\delta_{i,j+1})]/c_{0} + O(\kappa^{2})$.
The calculation in many respects is similar to the one in the
previous section, but a little easier. Specifically, we obtain

\begin{eqnarray}
  \Sigma_{Y}= n\big[\langle w^{2}\rangle\langle f^{2}\rangle
  - \langle w\rangle^{2}\langle f\rangle^{2}\big]
 \nonumber \\
  + 2(n-1)\kappa\langle w\rangle^{2} Var(f) 
\end{eqnarray}
and

\begin{eqnarray}
  Cov(Y,f_{i})= \langle w\rangle Var(f)
  \Big( 1 + \kappa[\delta_{i1}+\delta_{in}
 \nonumber \\
    +2(1-\delta_{i1})(1-\delta_{in})]\Big)
  + O(\kappa^{2}).
\end{eqnarray}
With the help of these relations we find

\begin{eqnarray}
  Cov(Y,\vec{f})^{T}\Sigma_{f}^{-1} Cov(Y,\vec{f})=
  \langle w\rangle^{2} Var(f)
  \nonumber \\
 \times\big[ n + 2(n-1)\kappa \big]
  + O(\kappa^{2}),
\end{eqnarray}
which leads to the Gaussian MI in the form

\begin{eqnarray}
\mbox{MI}(\vec{f},Y)_{g} \approx
\frac{1}{\ln4}\ln\Big( \Big[ 1 + \frac{\langle w\rangle^{2} Var(f)}
  {\langle f^{2}\rangle Var(w)} \Big]
  \nonumber \\
\times \Big[ 1 + \frac{2\kappa\langle w\rangle^{2} Var(f)}
  {[\langle f^{2}\rangle Var(w) + \langle w\rangle^{2} Var(f)]} + O(\kappa^{2}) \Big] \Big).
\end{eqnarray}
Comparing the above MI for Gaussian distributions with the corresponding MI for
lognormal in Eq. (32), we can notice significant differences, as
$\sigma_{f}^{2}, \sigma_{w}^{2}$ depend in a nonlinear way on $Var(f), Var(w)$.
Generally MI$_{ln}$ is greater than MI$_{g}$, and the discrepancy between
them grows with increasing presynaptic firing rate variability,
and with decreasing synaptic noise (Fig. 4). Additionally, MI$_{ln}$ diverges
for some finite $\sqrt{Var(f)}/\langle f\rangle$ for a fixed level of
$\sqrt{Var(w)}/\langle w\rangle$ (and vice versa), whereas MI$_{g}$ is
always finite for finite means and variances. 
However, in the case when variabilities in $f$ and $w$ are weak, i.e., for
$\sigma_{f}^{2}, \sigma_{w}^{2} \ll 1$, we see that the lognormal formula for MI
in Eq. (35) is exactly the same as the one in Eq. (39) for Gaussian MI
(note that in this limit $\langle f^{2}\rangle \approx \langle f\rangle^{2}$,
and $\kappa \mapsto \gamma$). This reflects a simple fact that for small
$\sigma_{f}^{2}, \sigma_{w}^{2}$ lognormal distributions can be approximated
by normal distributions \cite{romeo}.

\begin{figure}[h!]
    \centering
    \includegraphics[scale=0.62]{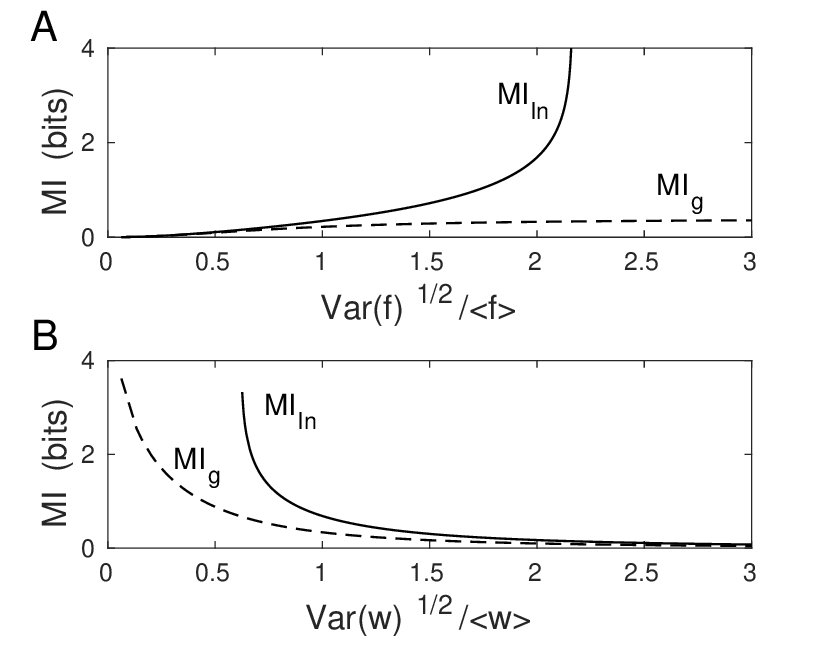}
    \caption{Neuronal MI as a function of presynaptic activity (A)
      and synaptic weights (B). Parameters used:
      (A) $Var(w)/\langle w\rangle^{2}= 1.4$ \cite{song};
      (B) $Var(f)/\langle f\rangle^{2}= 1.5$ \cite{hromadka}; and
    for both $n=8000$ \cite{braitenberg}. }
    \label{fig:enter-label}
\end{figure}

\subsection{\label{sec:levell3} Implications for information processing
  in cortical brain networks.}

The results shown in Fig. 4 indicate that high variability in neural activities
fosters information transfer between neurons. Morever, there seems to be a
fundamental difference between short-tailed and heavy-tailed distributions
of neural activities, with the latter having much bigger transmission impact
for a given ratio of standard deviation to the mean activity. This is also true
for the variability in synaptic weights: synapses with broad heavy-tailed
distributions generally provide higher mutual information than synapses with
bell-shaped distributions (Fig. 4B).

Our theoretical result, present in Fig. 4, and its interpretation are
compatible with experimental data coming from cortical networks
\cite{shew,fagerholm}. These papers demonstrated that diversity of brain
dynamics, i.e., activity patterns with broad heavy-tailed distributions
tend to improve information transfer between different cortical regions
\cite{shew,fagerholm}. More generally, these conclusions are also in line
with empirical evidence showing that neuronal representations in mammalian
brains are high-dimensional, meaning that neurons exhibit a diversity of
context dependent activities, which is important functionally \cite{fusi}.
Taken together, this may suggest that evolution prefers brains with
heterogeneous dynamics to optimize information processing
(e.g. \cite{buzsaki}).

\section{\label{sec:level4}Conclusions}

In this paper we used analytical expression for mutual information
between random vectors with lognormal distribution to obtain closed-form
expressions of MI for different networks of interacting elements with
specific covariance matrices, mostly sparse with a high degree of symmetry.
These formulas may be helpful in many practical applications in engineering
and biology. Additionally, we applied these results to information transfer
in neural networks of the mammalian cerebral cortex. Specifically, we derived
an analytical formula for MI of a neuron receiving many correlated synaptic
inputs that are lognormally distributed, and compared such MI with the case
when the total synaptic input is normally distributed. Interestingly, mutual
information in the first case of lognormal input can be significantly greater
than in the second case with Gaussian variables.

\section{Acknowledgments}

The work was supported by the Polish National Science Centre (NCN) grant number
2021/41/B/ST3/04300 (JK).

\vspace{0.5cm}

\noindent $^{*}$ Corresponding author: jkarbowski@mimuw.edu.pl

\appendix

\section{Inverse and determinant of tridiagonal matrix, and proof of Eq. (17).}
To compute  $Cov_{XY}^{T}\Sigma_{X}^{-1}Cov_{XY}$ in Eq. (17), we need
first to find the inverse of the tridiagonal matrix $\Sigma_{X}$. Let us
consider a general $n\times n$ tridiagonal matrix $A$ with all diagonal elements
equal to $t$, and off-diagonal elements all equal to $s$. Inverse elements of
$A$ are given by \cite{hu,usmani}

\begin{eqnarray}
(A^{-1})_{ij}= (-1)^{i+j}s^{j-i}\theta_{i-1}\phi_{j+1}/\theta_{n}
\end{eqnarray}
for $i \le j$, where $\theta_{n}= \det(A)$. Matrix $A^{-1}$ is symmetric,
and hence $(A^{-1})_{ij}= (A^{-1})_{ji}$. In Eq. (A1), $\theta_{i}$ and
$\phi_{i}$ are functions of $t$ and $s$. Additionally $\theta_{i}$ satisfy the
following recurrence relations

\begin{eqnarray}
\theta_{i}= t\theta_{i-1} - s^{2}\theta_{i-2}
\end{eqnarray}
with $\theta_{0}= 1$, and $\theta_{1}= t$. Similarly,
$\phi_{i}$ satisfy the following recurrence relations

\begin{eqnarray}
\phi_{i}= t\phi_{i+1} - s^{2}\phi_{i+2}
\end{eqnarray}
with $\phi_{n+1}= 1$, and $\phi_{n}= t$. 

Both of the recurrence relations, Eqs. (A2) and (A3), can be solved by
a standard substitution $\theta_{i}= r^{i}$, with unknown $r$. The solution
for $r$ is $r_{\pm}= \frac{1}{2}(t \pm \sqrt{t^{2}-4s})$. The specific solutions
for $\theta_{i}$ and $\phi_{i}$ are given by linear combinations of $r_{-}$ and
$r_{+}$, with coefficients dependent on the boundary conditions in Eqs. (A2)
and (A3). It is easy to show that

\begin{eqnarray}
 \theta_{i}(t,s) = 
    \frac{\Big[ \big(t + \sqrt{t^2-4s^2}\big)^{i+1} 
       - \big(t - \sqrt{t^2-4s^2}\big)^{i+1} \Big]}
      {2^{i+1}\sqrt{t^2-4s^2}}
 \nonumber \\
\end{eqnarray}
and $\phi_{i}= \theta_{n-i+1}$, for $i=1,2,...,n$. The functions $\theta_{i}(t,s)$
can be interpreted as determinants of reduced $i\times i$ matrices generated
from the original matrix $A$, with all diagonal elements $t$ and all off-diagonal
elements $s$.

Now we can determine $Cov_{XY}^{T}\Sigma_{X}^{-1}Cov_{XY}$ in Eq. (17).
We have

\begin{eqnarray*}
\sum_{i,j=1}^{n}(Cov_{XY}^{T})_{i}(\Sigma_{X}^{-1})_{ij}(Cov_{XY})_{j} =
a^{2}\sum_{i,j=1}^{n} (\Sigma_{X}^{-1})_{ij}
 \\
 = a^{2}\Big(  \sum_{i=1}^{n} (\Sigma_{X}^{-1})_{ii}
 + 2\sum_{i=1}^{n-1}\sum_{j=i+1}^{n} (\Sigma_{X}^{-1})_{ij} \Big),
\end{eqnarray*}
where the symmetric matrix $\Sigma_{X}^{-1}$ has the following elements

\begin{eqnarray}
  (\Sigma_{X}^{-1})_{ij}= \frac{(-1)^{i+j}\gamma^{j-i}}{\det(\Sigma_{X})}
  \theta_{i-1}\theta_{n-j},
\end{eqnarray}
for $i \le j$. These sums can be executed, and after lengthly computations
one can obtain Eq. (17) in the main text.


\end{document}